\documentstyle[12pt]{article}                                                   
\newcommand{\bibi}{\bibitem}                                                    
\newcommand{\etal}{\it {et al.}}                                                
\newcommand{\prl}{\it Phys. Rev. Lett.}                                         
                                             
\newcommand{\prb}{\it Phys. Rev. B}

\newcommand{\half}{\frac {1}{2}}                                                
                                               
\newcommand{\beq}{\begin{equation}}                                             
\newcommand{\eeq}{\end{equation}\noindent}                                      
                                                                                
\newcommand{\beqr}{\begin{eqnarray}}                                            
\newcommand{\eeqr}{\end{eqnarray}\noindent}

\newcommand{\vk}{{\bf k}}                                                       
\newcommand{\vq}{{\bf q}}                                                       
\newcommand{\ku}{\vk\uparrow}                                                   
\newcommand{\kd}{-\vk\downarrow}                                                
\newcommand{\kqd}{-\vk + \vq\downarrow}                                         
\newcommand{\vkp}{{\bf k}^{\prime}}                                             
\newcommand{\cd}{c^{\dag}}                                                      
\newcommand{\uw}{\uparrow}                                                      
\newcommand{\dw}{\downarrow}                                                    
\newcommand{\udw}{\uparrow\downarrow}                                           
\newcommand{\noin}{\noindent}                                                   
                                                       
\begin{document}                                                                
\title {Spin Gap and Superconductivity in the                                   
Interlayer Pair Tunneling Model}                                                
                                                                                
\author{Sanjoy K. Sarker \\                                                     
Department of Physics and Astronomy \\                                          
The University of Alabama, Tuscaloosa, AL 35487 \\                              
and\\                                                                           
Philip W. Anderson\\                                                            
Joseph Henry Laboratories of Physics\\                                          
Princeton University, Princeton, NJ 08544\\}                                    
\date{}                                                                         
\maketitle                                                                      
\begin{abstract}                                                                
A simple interlayer pair tunneling model is solved exactly.                     
We find that in the normal state                                                
the spin-$\half$ particle and hole excitations are gapped.                      
But the state is an unusual metal, characterized by novel fermionic,            
spin-zero and charge $\pm 2e$ gapless excitations that exist about new          
type of Fermi surfaces. The model is consistent with a number of unusual        
properties of underdoped cuprates.                                              
Superconductivity is induced by an additional intra-layer pairing               
interaction which opens a gap                                                   
in the charge spectrum. The symmetry of the order parameter                     
is in general different from the symmetry of the                                
single-electron (or hole) gap.                                                  
The former can be a d-wave, while the the latter is more complex.               
                                                                                
\end{abstract}                                                                  
                                                                                
\vspace{0.5 in}                                                                 
                                                                                
\pagebreak

In underdoped bilayer cuprate superconductors there appears to be a gap,        
known as the spin gap, in the normal-state spectrum. For example, below         
about 150 K, magnetic susceptibility $\chi$ and NMR relaxation                  
rate decrease rapidly \cite{tak}. Yet there is apparently no charge             
gap since the in-plane conductivity is metallic.                                
Recent photoemission experiments                                                
further show that only a fraction of the                                        
the Fermi surface near $(\pi,0)$ and symmetric points seems to                  
be gapped \cite{mar}. On continuity grounds, Fermi or Luttinger                 
liquids are not expected to show such a behavior, suggesting that these         
systems may belong to a different universality class. Since the spin-gap        
has been observed only in bilayer systems, it has been suggested that           
interlayer interactions may be responsible \cite{mil}.                          
An interesting possibility is interlayer pair tunneling                         
which was initially considered as a mechanism for $T_c$                         
enhancement \cite{and1}.                                                        
The underlying assumption is that single electrons can not                      
hop coherently between copper-oxide layers so that second order                 
hopping processes such as pair tunneling become important.                      
                                                                                
Recently Chakravarty {\etal} have studied a simple model of pair                
tunneling between two layers and found considerable enhancement of              
$T_c$ within a mean-field approximation \cite{sud}. Physics within              
each layer is described by a Bardeen, Cooper and Schrieffer (BCS)               
type reduced Hamiltonian, and tunneling is also                                 
confined to the reduced subspace of pairs with                                  
zero total momentum. The mean-field state does not have a spin gap.             
But, as shown by                                                                
Anderson, when the tunneling interaction is treated exactly                     
the normal-state $\chi$ exhibits spin-gap behavior \cite{and2}.                 
In this paper we show that, although the model is too simple to provide         
a detailed description of the cuprates, it has a number interesting             
properties. We find that in the normal state spin-$\half$ electron and          
hole excitations acquire a gap. The Fermi surface appears partially             
gapped at finite $T$ (experimental situation).                                  
An important property is that there are gapless                                 
spin-$0$, charge $\pm 2e$ excitations which are {\em fermionic}                 
and have their own Fermi surface, and which dominate the low-$T$ physics.       
The system is thus metallic, but not a Fermi liquid in the usual sense.         
To induce superconductivity we add an intralayer                                
attractive interaction which opens a gap in the charge spectrum.                
Therefore the symmetry of the charge gap                                        
(and hence the order parameter) is determined by the attractive                 
interaction and is in general different from the                                
symmetry of the one particle (i.e., spin-) gap.                                 
                                                                                
We consider the following Hamiltonian for two layers                            
\beq H = \sum _{i\vk\sigma}\epsilon (\vk)\cd_{i,\vk\sigma}c_{i\vk\sigma}        
~ - ~ \sum_{\vk,\vkp} V_{\vk\vkp}                                               
\cd_{i\ku}\cd_{i,\kd}c_{i,-\vkp \downarrow}c_{i,\vkp \uparrow}                  
~~-~\sum _{\vk}                                                                 
 T_J(\vk)\lbrack \cd_{1,\ku}\cd_{1,\kd}c_{2,\kd}c_{2,\ku}~~+~~h.c.              
 \rbrack. \eeq                                                                  
Here $c_{i,\vk\sigma}$ destroys an electron carrying a                          
(two-dimensional) wavevector                                                    
$\vk$, spin $\sigma = \uparrow, \downarrow$ in layer $i=1,2$.                   
The first term describes an electron band                                       
of energy $\epsilon (\vk)$ for each layer. The second term                      
is an in-plane attractive interaction and the third term                        
decribes tunneling. Like the BCS Hamiltonian                                    
only pairs of zero total momentum are considered in the interaction             
terms so that tunneling matrix element $T_J(\vk)$ is diagonal in $\vk$.         
The assumption is that $T_J(\vk)$ is O(1). This is different from               
more traditional interactions such as $V_{\vk\vkp}$ (or tunneling               
terms with nonzero momentum) which scale inversely as the                       
volume of the system                                                            
and hence does not contribute for $\vk = \vkp$.                                 
It is precisely the diagonal nature of $T_J(\vk)$ that                          
causes the enhancement of $T_c$ and the spin gap. Although the problem          
can be solved for any $T_J(\vk)$, for definiteness we will take                 
\beq T_J(\vk) = \frac{t_J}{16}(\cos k_x - \cos k_y)^4. \eeq                     
This is the same form used in the mean-field                                    
analysis \cite{sud}. Our focus will be on the universal features and            
as such we will take a simple cosine band:                                      
$\epsilon (\vk) = -2t\lbrack \cos k_xa + \cos k_ya\rbrack.$                     
                                                                                
We first ignore $V$ and solve the normal state problem.                         
Then the Hamiltonian can be diagonalized exactly by diagonalizing               
each $\vk$ subspace separately \cite{str}.                                      
For a given $\vk$, we need to consider                                          
four single-electron states: $(1,\ku),(1,\kd), (2,\ku)$ and $(2,\kd)$.          
It is convenient to diagonalize ${\cal H_{\vk}}= H_{\vk} -                      
\mu N_{\vk}$, where $N_{\vk}$ is the number operator and                        
$\mu$ is the chemical potential.  Let $\xi_{\vk} ~= \epsilon (\vk)~-            
~\mu$. There are 16 many-body states which can                                  
be grouped according to the total electron number $N_{\vk}$ and                 
z-component of total spin $S_z$ which are both conserved.                       
In the absence of tunneling, an eigenstate has energy $N_{\vk}\xi_{\vk}$,       
with $N_{\vk} = 0,1,2,3,4$. We represent such a state by                        
$|a~~~~b>$, where $a$ ($b$) stands for one of                                   
the four states in the first (second) layer: $0,~ \uw,                          
\dw,~ \uw\dw$. Tunneling only connects two of these:                            
$|\uw\dw~~~~0>$ and $|0~~~~\uw\dw>$, with $N_{\vk} = 2$ and                     
$S_z = 0$. Diagonalizing the 2 by 2 matrix                                      
leads to the following two states:                                              
\beq |2 \pm> = \frac{1}{\sqrt 2} \lbrack |\uw\dw ~~~~0>~ \pm~                   
 |0~~~~\udw>\rbrack, \eeq                                                       
with energies $2\xi_{\vk} ~\mp ~T_J(\vk)$, respectively. The remaining          
fourteen states are unaffected. Note that since $N_{\vk}$ is conserved,         
there is no long-range order, i.e, there is no superconductivity in             
the absence of $V$.                                                             
                                                                                
\medskip                                                                        
                                                                                
\noindent{\underbar{Ground State:}}                                             
The ground state of the full Hamiltonian is constructed                         
by selecting the lowest energy state of $\cal H_{\vk}$ for each $\vk$.          
The relevant states are the (i) the four-electron state $|4> \equiv             
|\udw~~~~\udw>$ of energy $4\xi_{\vk}$; (ii) the                                
zero-electron state $|0> \equiv |0~~~~0>$ of energy $0$ and (iii)               
the two-electron state $|2 +>$ of energy $2\xi_{\vk} - T_J(\vk)$.               
For $T_J = 0$, states with different $N_{\vk}$ cross at the Fermi level         
which is at $\xi _{\vk} = 0$.                                                   
So one must pick the four-electron state for $\xi_{\vk} < 0$ and                
the zero-electron state for $\xi_{\vk} > 0$.                                    
                                                                                
For nonzero $T_J(\vk)$, the four-electron state still has the lowest            
energy for $\xi_{\vk} < - \half T_J(\vk)$ and the zero-electron state           
for $\xi_{\vk} >  \half T_J(\vk)$. But in the region                            
$- \half T_J (\vk) < \xi_{\vk} < \half T_J(\vk)$, the two-electron              
state $|2 +>$  has the lowest                                                   
energy. We will call this the two-electron region.                              
{\em Therefore the original Fermi surface has disappeared,                      
except at points where                                                          
the line $T_J(\vk) = 0$ intersects the Fermi surface                            
($\xi _{\vk} = 0$)}.                                                            
In its place, two new \lq\lq Fermi" surfaces have appeared.                     
Surface I is at $\xi _{\vk} = - \half T_J(\vk)$ and                             
separates the four-electron states from the two-electron states. Surface        
II appears at $\xi_{\vk} = \half T_J(\vk)$ and separates                        
the two-electron states from the zero-electron states. As shown in Fig 1,       
the particle number $N_{\vk}$ has discontinuities at the new                    
Fermi surfaces whose topolgy depends on the form of $T_J(\vk)$.                 
In the quartic case, $T_J(\vk)$ is largest near $(0,\pm\pi)$                    
and $(\pm\pi,0)$, but it vanishes along $k_x = k_y$,                            
so that the original Fermi surface survives at four points.

\medskip                                                                        
\noin {\underbar {Excitations:}}                                                
In the absence of tunneling, the elementary excitations are                     
the usual holes, created by picking                                             
a three-electron state in the four-electron region and particles,               
created by picking a one-electron state in the zero-electron                    
region. These are gapless (energy $|\xi _{\vk}|$),                              
carry charge $\pm e$ and spin $\half$.                                          
                                                                                
(i) For $T_J(\vk)> 0$, the hole and particle excitations described above        
exist for $\xi _{\vk} < - \half T_J(\vk)$                                       
and  $\xi _{\vk}> \half T_J(\vk)$, respectively.                                
Two additional spin-$\half$ excitations exist in                                
the two-electron region: a particle of energy $\xi _{\vk} + T_J(\vk)$           
and a hole of energy $-\xi _{\vk}+ T_J(\vk)$. {\em All                          
four spin-$\half$ excitations are gapped}. Therefore, magnetic                  
susceptibility                                                                  
$\chi$ is zero at $T = 0$. The \lq\lq spin gap" is thus a                       
natural consequence of tunneling, will be observed in                           
photoemission experiments. We define a gap function                             
$\Delta _{spin}(\vk)$ by the value of the excitation energy                     
at the new Fermi surface. This equals $\half T_J(\vk)$,                         
exactly the distance between the old and new Fermi surfaces.                    
                                                                                
(ii) Near the new Fermi surfaces there are additional                           
{\em gapless} hole- and particle-type excitations.                              
However, they have spin zero and charge {$\pm 2e$}. Near surface I, a           
particle is created by selecting the four-electron state in the                 
two-electron region and a hole is created by selecting a two-electron           
state in the four-electron region. These have energy                            
$\pm(2\xi _{\vk}+ T_J(\vk))$ which vanish at surface I                          
($\xi _{\vk} = - \half T_J(\vk)$), and can not be created                       
from two spin-$\half$ excitations. Similar spin-zero, charge-two                
particle and                                                                    
hole excitations also exist near Fermi surface II with energies                 
$\pm(2\xi _{\vk} ~ - ~ T_J(\vk))$.                                              
                                                                                
In short, while there is a spin gap, there is no {\em charge gap}.              
The charge excitations                                                          
can not be viewed as \lq preformed' Cooper pairs since, as shown                
below, they behave like fermions,                                               
at least as far as thermodynamic properties are concerned.                      
The partition function for the $\vk$ subspace is given by                       
\beq Z(\vk) = Z_0(\vk)~+~ Z_1(\vk) ~=~                                          
 (1~+~e^{-\beta\xi_{\vk}})^4 +                                                  
 2\lbrack \cosh(\beta T_J(\vk))~-~1\rbrack                                      
e^{-2\beta\xi_{\vk}},\eeq                                                       
where $Z_0$ is the partition function for $T_J = 0$, and $\beta = 1/kT$.        
Thermodynamic properties are largely determined by the behavior                 
of the two factors $Z_0/Z$ and $Z_1/Z$ as a function of $T$.                    
At high temperatures, $kT >> T_J$, $Z_1 \rightarrow 0$ and we                   
recover the noninteracting result. But at low temperatures such                 
that $\beta T_J >> 1$ and $\beta|\xi| >> 1$, the two factors look               
like Fermi functions. Thus for $\xi < 0$                                        
\beq Z_0/Z \sim  f(2\xi ~+~ T_J), ~~~~ Z_1/Z                                    
 \sim  (1 - f(2\xi ~+~ T_J)).\eeq                                               
And for $\xi > 0$                                                               
\beq Z_0/Z \sim (1 ~-~ f(2\xi ~-~ T_J)), ~~~~ Z_1/Z \sim                        
f(2\xi ~-~ T_J).\eeq                                                            
Note that these functions have step discontinuities at the new                  
Fermi surfaces, and the energy argument correspond to the                       
charge excitations. Of course, far from the Fermi surfaces they                 
lose their fermionic character.                                                 
                                                                                
We will be interested in the case for which the noninteracting                  
Fermi energy $\epsilon _{F0} >> t_J, kT$.                                       
As shown in Fig. 1, for $kT << T_J$, the new Fermi surfaces and                 
the fermionic character of the charge excitations are                           
evident in the particle number $n_{\vk} \equiv N_{\vk}/4$,                      
which is given by                                                               
\beq n_{\vk} = \frac{Z_0(\vk)}{Z(\vk)}f(\xi_{\vk}) +                            
\frac{Z_1(\vk)}{2Z(\vk)}.\eeq                                                   
With increasing $T$, the charge excitations break up and density                
of spin-$\half$ excitations increases, and                                      
at $kT \sim T_J$, the new Fermi surfaces have effectively                       
disappeared so that $n_{\vk}$ looks like                                        
the Fermi distribution function for the noninteracting system.                  
Therefore the gap $\Delta _{spin}(\vk)$ will be seen only on those              
parts of the Fermi surface where $T_J(\vk)$ is large compared to                
$kT$. For the quartic case, $T_J(\vk)$ vanishes at four points                  
along the (1,1) direction, and is quite small (because of the                   
fourth power) except in the neighborhood of                                     
$(\pm\pi,0)$ and $(0,\pm\pi)$, where the gap is largest. Hence, for             
$T > 0$, parts of the Fermi surface along (1,1)                                 
and nearby directions will appear to be ungapped.                               
For a fixed $T$, the gap decreases with doping,                                 
as the old Fermi surface moves away from the maximal points. For                
fixed density it effectively disappears for $kT > t_J$.                         
These results are in qualitative agreement with                                 
recent photoemission measurements \cite{mar}.                                   
                                                                                
The state is clearly not the usual Fermi liquid. This is true for               
arbitrarily small $t_J$, and all densities and spatial dimensions               
(there is no quantum phase transition). Moreover, there is no broken            
symmetry. For $kT << t_J$, charge fermions dominate and make the system a       
degenerate metal. With increasing $T$                                           
these give way to spin-$\half$ excitations, and eventually at                   
$kT \sim t_J$ there is a crossover to an ordinary Fermi liquid.                 
                                                                                
The susceptibility is determined by the spin-$\half$                            
excitations, and is given by \cite{and2}                                        
\beq \chi = \frac{\mu_0^2}{2N} \sum _{\vk} \frac{Z_0(\vk)}{Z(\vk)}              
\left[\frac{\beta e^{\beta\xi _{\vk}}}{(e^{\beta\xi _{\vk}}~ + ~1)^2}           
\right], \eeq                                                                   
where $\mu _0 = e\hbar/2mc$. The quantity in the bracket is                     
localized at the old Fermi surface ($\xi = 0$) and has width $kT$.              
For $kT < T_J$, its overlap with the factor $Z_0/Z$ is exponentially            
small. Thus if $T_J(\vk)$ has                                                   
no zeroes at the Fermi level (finite gap), then $\chi$ vanishes                 
exponentially with $T$.                                                         
Suppose $T_J(\vk)$ is a constant = $2\Delta _{spin}$.                           
Then, for $kT << T_J << \epsilon _{F0}$,  we find that                          
$\chi/\chi _0 \approx (kT/\Delta _{spin})exp(-2\Delta _{spin}/kT)$,             
where $\chi _0 = \half\rho _0\mu _0^2$ is the zero-temperature                  
susceptibility for the noninteracting system and                                
$\rho _0$ is the corresponding density of states at the Fermi level.            
For $kT >> t_J$, on the other hand, we recover the noninteracting               
result: $\chi /\chi_0 \approx 1 - \frac{2}{15}(\Delta _{spin}/kT)^2$.           
                                                                                
If the gap function has zeroes one expects a power law. The power               
depends on the symmetry of $\Delta _{spin}(\vk)$.                               
For the quartic case, $T_J(\vk)$ (and hence the gap)                            
vanishes at the Fermi points $\vk _0$ as $T_J(\vk) \propto                      
(q_x - q_y)^4$,                                                                 
where ${\bf q} = \vk - \vk _0$. Then we find that $\chi$ vanishes               
as $\chi \sim T^{1/4}$. More generally $\chi \sim T^{1/p}$, if                  
the spin gap vanishes with a power $p$.                                         
Fig 2. shows the temperature dependence of $\chi$, calculated                   
numerically, for several values of the electron density $n$.                    
We see that $\chi$ deviates sharply from $\chi _0$ below a                      
temperature $T_{sus}$ and then decreases slowly                                 
as $T^{1/4}$. The scale $T_{sus}$ decreases with                                
decreasing $n$.                                                                 
                                                                                
For $T \rightarrow 0$ charge-fermions dominate. For example,                    
specific heat per site $C(T) \equiv T\gamma (T)$                                
vanishes linearly with $T$. Fig. 3 shows the temperature dependence of          
$\gamma (T)/\gamma _0$, where $\gamma _0$ is the coefficient for the            
noninteracting system. We find that at low $T$,                                 
$\gamma (T) /\gamma _0 \approx 1/4 + aT^{1/4}$.                                 
The constant term $1/4$ arises because, for charge fermions,                    
the spin-degeneracy is reduced by a factor of $2$ and their                     
Fermi energy is incresed by a factor of 2,                                      
to leading order in $t_J/\epsilon _{F0}$. The $T^{1/4}$ is again                
the contribution from the spin-$\half$ excitations.                             
As shown  in Fig. 3, $\gamma (T)$                                               
goes through broad maximum at $T = T_{spht}$, and                               
at large $T$ approaches the nonointeracting value $\gamma _0$.                  
Thus the Wilson ratio ($\propto C/T\chi$) is temperature dependent              
and becomes infinite at $T = 0$. The scale $T_{spht}$ is in general             
much smaller than $T_{sus}$.                                                    
                                                                                
{\underbar{Superconductivity:}}                                                 
To induce superconductivity we now add the in-plane                             
interaction $V_{\vk,\vkp}$. It is sufficient to treat this term by              
mean-field approximation since it is restricted to the reduced subspace         
of zero-momentum pairs. But $T_J$ is again treated exactly. Let                 
$b_{\vk} \equiv  <\cd_{i\ku}\cd_{i,\kd}>$ be the order-parameter which          
we take to be real and independent of the layer index.                          
Then the mean-field pairing Hamiltonian is                                      
\beq H_{super,MF} =  - \sum _{i,\vk}                                            
\Delta_{\vk}\left[\cd_{i,\ku}\cd_{i,\kd}                                        
~~+~~c_{i,\kd}c_{i,\ku}\right],  \eeq                                           
where                                                                           
$\Delta_{\vk} = \sum _{\vkp} V_{\vk\vkp}b_{\vkp}$.                              
The full Hamiltonian can again be diagonalized for each                         
$\vk$ separately. The total $S_z$ is still a                                    
good quantum number for each $\vk$, but $N_{\vk}$ is no longer                  
conserved. States with $S_z \ne 0$ are not affected by tunneling,               
and therefore are the same as in the BCS theory.                                
In particular, the $S_z = \pm 1/2$ subspaces                                    
split into two                                                                  
manifolds of energy $2\xi _{\vk} \pm R_{\vk}$, where                            
$R_{\vk}~=~\left(\xi _{\vk}^2 ~+~ \Delta_{\vk}^2\right)^{\half}$.               
                                                                                
The lowest-energy eigenstate belongs to the same                                
$S_z = 0$ subspace as before, the subspace consisting of                        
the three states $|0>$, $|2 +>$ and $|4>$.                                      
These are now coupled by the pairing                                            
Hamiltonian leading to the following 3 by 3 matrix:                             
\[ \left| \begin{array}{ccc}                                                    
   0 & - \sqrt 2  \Delta_{\vk} & 0 \\                                           
-\sqrt 2 \Delta_{\vk} & 2\xi _{\vk} - T_J(\vk) &  -\sqrt 2\Delta_{\vk} \\       
   0 & - \sqrt 2 \Delta_{\vk}  & 4\xi _{\vk}                                    
\end{array}\right|.\]                                                           
The eigenvalues are given by                                                    
\beq \Lambda _m = 2\xi _{\vk} - \frac{1}{3} T_J(\vk) +                          
2\left(\frac{d}{3}\right)^{\half} \cos \lbrack \frac{1}{3}(2\pi m+              
\phi)\rbrack, \eeq                                                              
where $m = 1,2,3$, and                                                          
$\tan \phi = - \left(\frac{4d^3}{27} - c^2\right)^{\half}/c$,                   
$d = 4\xi_{\vk}^2 + 4\Delta_{\vk}^2 + \frac{1}{3} T_J^2(\vk)$ and               
$c = \frac{2}{3}T_J(\vk)\left[2\Delta_{\vk}^2 + \frac{1}{9}T_J^2(\vk)           
- 4\xi_{\vk}^2\right].$                                                         
                                                                                
\medskip                                                                        
                                                                                
\noin{a. \underbar{Charge Gap:}}  For $T_J =0$ we have the usual                
BCS problem. The                                                                
spin-$\half$ particles and holes acquire a gap $\Delta _{\vk}$                  
at the original Fermi surface. For $T_J \ne 0$,                                 
these are already gapped in the normal state.                                   
The main source of the condensation energy is                                   
a gap that opens up in the charge fermion spectrum.                             
For example, $|0>$ and $|2 +>$ are degenerate with energy 0.                    
at the (new) right Fermi surface ($\xi = T_J/2$).                               
This is split by pairing, giving a charge gap which has the                     
form $\Delta _{charge}(\vk) =                                                   
\Delta _{\vk}f_1(\frac{\Delta _(\vk)}{2T_J(\vk)})$,                             
where $f_1(x)$ is                                                               
even in $x$. For $\Delta _{\vk}$ small compared to $2T_J(\vk)$,                 
we find that $\Delta _{charge}(\vk) \approx 2\sqrt 2 \Delta _{\vk}$.            
The superconducting $T_c$ is enhanced relative to the BCS value                 
in part                                                                         
because $\Delta _{charge}(\vk)$ is generally larger than $\Delta _{\vk}$.       
                                                                                
\noin{b.~ \underbar{Symmetry of the Order Parameter:}}                          
The order parameter $b_{\vk} = <c _{1\ku}c_{1\kd}>$                             
is an {\em odd} function of $\Delta _{\vk}$  since                              
a change of sign $\Delta _{\vk} \rightarrow -\Delta _{\vk}$ is                  
undone by a gauge transformation  $c_{\ku} \rightarrow                          
- c_{\ku}$ which changes the sign of $b_{\vk}$.                                 
At $T = 0$ it has the form                                                      
\beq b_{\vk} = \Delta _{\vk}                                                    
\left[\frac{(4\xi - \Lambda (\vk))^{-1} - \Lambda ^{-1}(\vk)}                   
{1 + 2\Delta _{\vk}^2\left(\Lambda  (\vk)^{-2}                                  
+ (4\xi -\Lambda  (\vk))^{-2}\right)}\right], \eeq                              
where $\Lambda (\vk)$ is the lowest eigenvalue of the 3 by 3 matrix             
and is even in $\Delta _{\vk}$. The symmetry of the order parameter             
(and of the charge gap) is                                                      
thus determined by the symmetry of the in-plane gap function                    
$\Delta_{\vk}$, which is determined by our choice of $V_{\vk,\vkp}$.            
In particular it could have d$_{x^2-y^2}$ symmetry.                             
On the other hand, the gap in the spin-$\half$                                  
electron and hole excitations --- the analog of the                             
BCS gap --- does not have                                                       
the same symmetry since it now depends both on $T_J(\vk)$ and                   
$\Delta _{\vk}$. In other words, experiments                                    
that probe the order parameter (e.g., Josephson tunneling) could                
find a d-wave symmetry, but experiments in which single electrons               
are added or removed (e.g., photoemission)                                      
would see a more complex gap structure.

\medskip                                                                        
                                                                                
A number of other results can be obtained                                       
independently of the origin or                                                  
details of $V_{\vk,\vkp}$.                                                      
As in the BCS model, $V_{\vk,\vkp}$ is presumed to be                           
appreciable near the original Fermi surface (at $\xi_{\vk}                      
 = 0$), i.e., for $\xi _{\vk}, \xi _{\vkp}                                      
< \omega _0$, where $\omega _0$ is                                              
the analog of the Debye energy.                                                 
Consequently $\Delta _{\vk}$ decreases with distance                            
from $\xi _{\vk} = 0$. But the charge gap                                       
scales with the value of $\Delta _{\vk}$ at the new Fermi surface               
which is at a distance $\half T_J(\vk) = \Delta _{spin}(\vk)$.                  
It follows that                                                                 
$\Delta _{charge}(\vk)$ is small where $\Delta _{spin}(\vk)$                    
is large (e.g., near $\pm\pi,0)$. This implies that close to half               
filling (underdoped regime) charge gaps are small, and increases                
with doping (decreasing $n$). Therefore, in this region,                        
(1) $T_c$ also increases with doping, as observed.                              
(2) For fixed $T << T_c$, density of charge                                     
excitations {\em decreases} with doping. Consquently, electronic                
contribution to specific heat, thermal conductivity etc                         
would decrease with increasing doping. (3) Since $T_c$ increases and            
$T_{sus}$ decreases with doping, the temperature range over which               
the spin gap is seen decreases with increasing doping and eventually            
disappears.                                                                     
                                                                                
In conclusion, the interlayer pair tunneling model provides                     
a simple example of the break down of Landau theory and                         
can account for a number of unusual properties of the underdoped                
cuprates. Since the effect is caused by a                                       
$\vk$-diagonal tunneling interaction one may ask whether such a                 
restriction is physically justified.                                            
For example, a more general tunneling Hamiltonian which results from            
second order hopping processes would have the form                              
\beq H_J = \sum _{\vk,\vq}                                                      
 T_J(\vk,\vq)\lbrack \cd_{1,\ku}\cd_{1,\kqd}c_{2,\kqd}c_{2,\ku}~~               
 +~~h.c.\rbrack, \eeq                                                           
where $\vq$ is the total momentum of the pair. Now, extensitivity               
of free energy requires that $\sum _{\vk,\vq} T_J(\vk,\vq)$ scales with         
the volume $V$, so that $T_J(\vk\vq) \propto V^{-1}$. But Hamiltonian           
(1) is obitained if only $\vq =0$ terms are kept. This is not                   
thermodynamically significant unless one                                        
postulates that $T_J(\vk,\vq = 0)$ is O(1), whereas                             
terms with $\vq \ne 0$ scale as $V^{-1}$ and hence are similar to               
other interaction processes. Note that this                                     
implies the existence of something akin to a                                    
condensate which favors zero-momentum pair tunneling.                           
Phenomenologically, if the diagonal terms are responsible for                   
the spin-gap state, then thermodynamic considerations                           
alone guarantee that they are fundamentally different from other                
interaction processes. It then makes sense to start with diagonal               
terms only, and treat the remainder as fluctuatuations.                         
Generalization to many layers is, in principle,                                 
straightforward. Indeed Baskaran has shown that, if $V$ is ignored, the         
model can be mapped onto a solvable quantum spin-chain model                    
along the c-axis                                                                
for each $\vk$, provided the Hamiltonian is restricted to the                   
$N_{\vk} = even$ (i.e., charge) subspace only \cite{bas}.                       
Then it is not clear what                                                       
happens to the spin gap. Unfortunately,                                         
when $V$ is included the spin-chain model is no longer solvable.                
                                                                                
We thank Tin Lun Ho and G. Baskaran for valuable comments.                      
This work was supported in part by the National Science Foundation,             
DMR-9104873 (P.W.A).                                                            
                                                                                
                                                           
                                                                                
\centerline{Figure Captions}                                                    
                                                                                
Fig. 1. Particle density $N_{\vk}$ along the $(\pi,0)$ direction.               
The intermediate                                                                
step with $N_{vk} = 2$ is the two-electron region.                              
At low $T =0.1t_J$ ($k_B =1$) there are two Fermi functions representing        
charge $\pm 2e$ excitations. But at $T = 0.5t_J$, there is only                 
one Function corresponding to the nontinteracting FS (which is                  
where the three lines intersect). Bandwidth = $40 t_J$.                         
                                                                                
Fig 2. (a) Magnetic susceptibility $\chi/\chi _0$                               
as a function of scaled temperature $t/t_J$, for four values of                 
electron density $n$. Note that characteristic temperature decreases            
with decreaing $n$. (b) At low $T$, $\chi$ vanishes as $T^{1/4}$,               
as shown in the log-log plot. The slope of the line is 1/4.                     
                                                                                
Fig 3. (a) Scaled specific heat cefficient                                      
$\gamma _(T)/\gamma _0 ~~~ = C(T)/C_0(T)$ as a function of                      
of scaled temperature $T/t_J$.                                                  
The broad maximum appears at a temperature which is much lower than             
temperature at which susceptibility drops (see Fig 2a) and may not be           
seen if superconductivity intervenes. (b) The same quantity on a                
log-log plate showing that it approaches a constant (=1/4 +                     
corrections) as $T \rightarrow 0$.

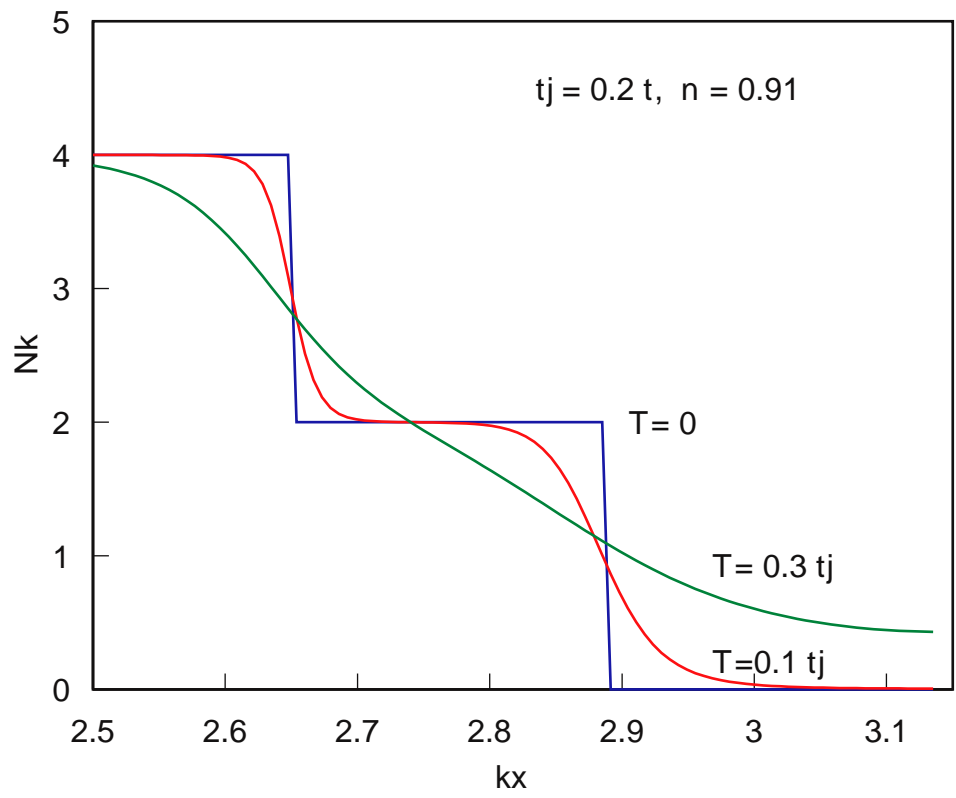
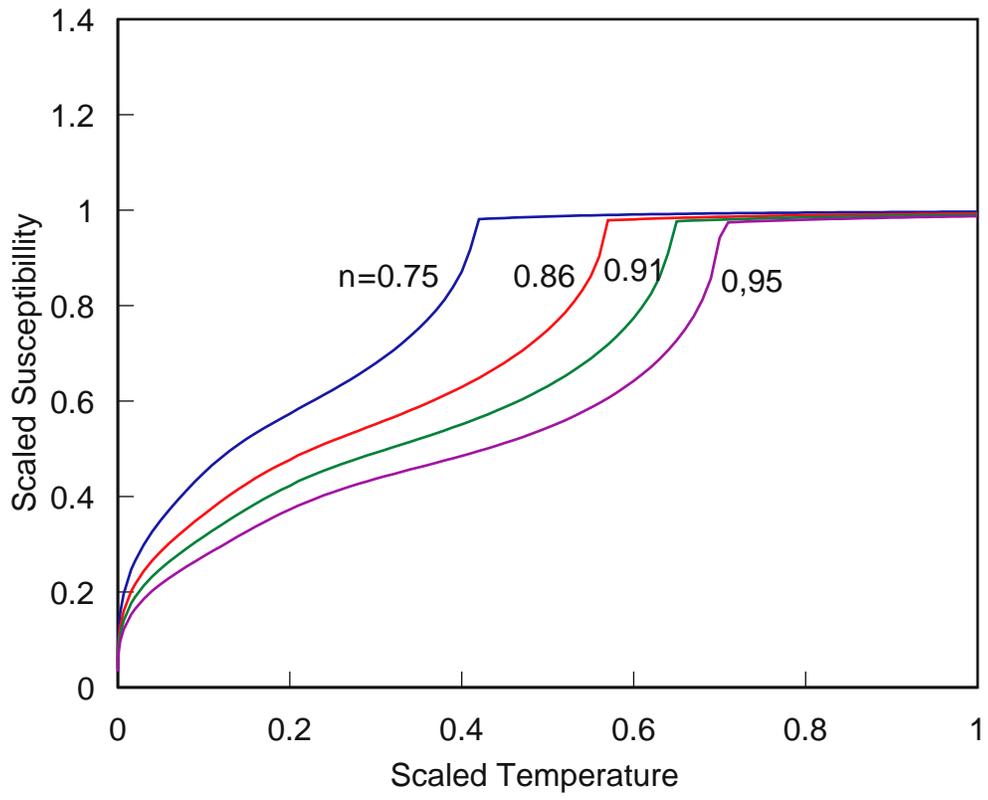
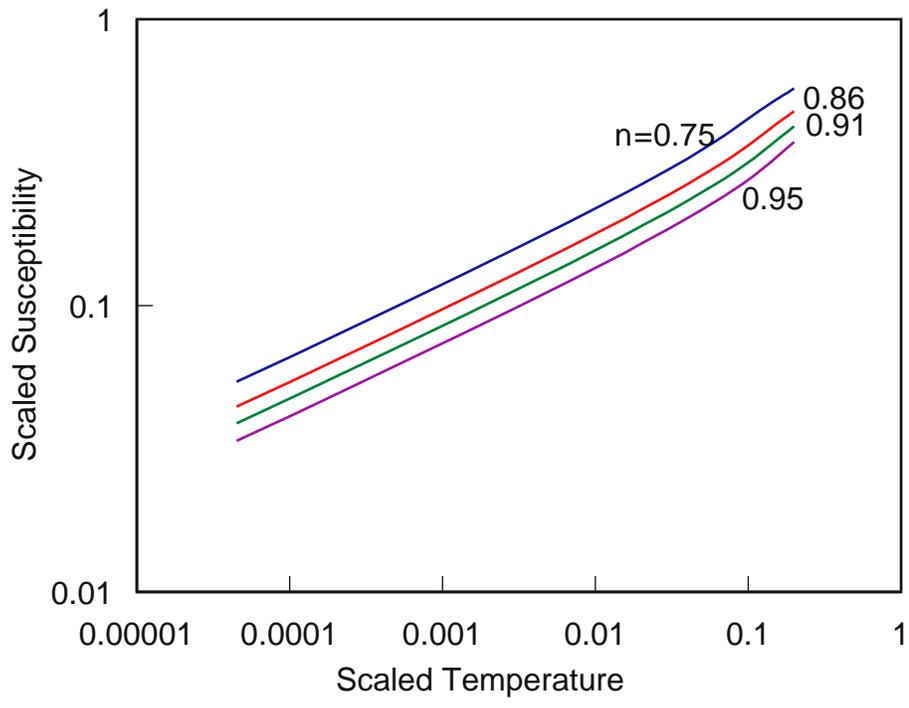
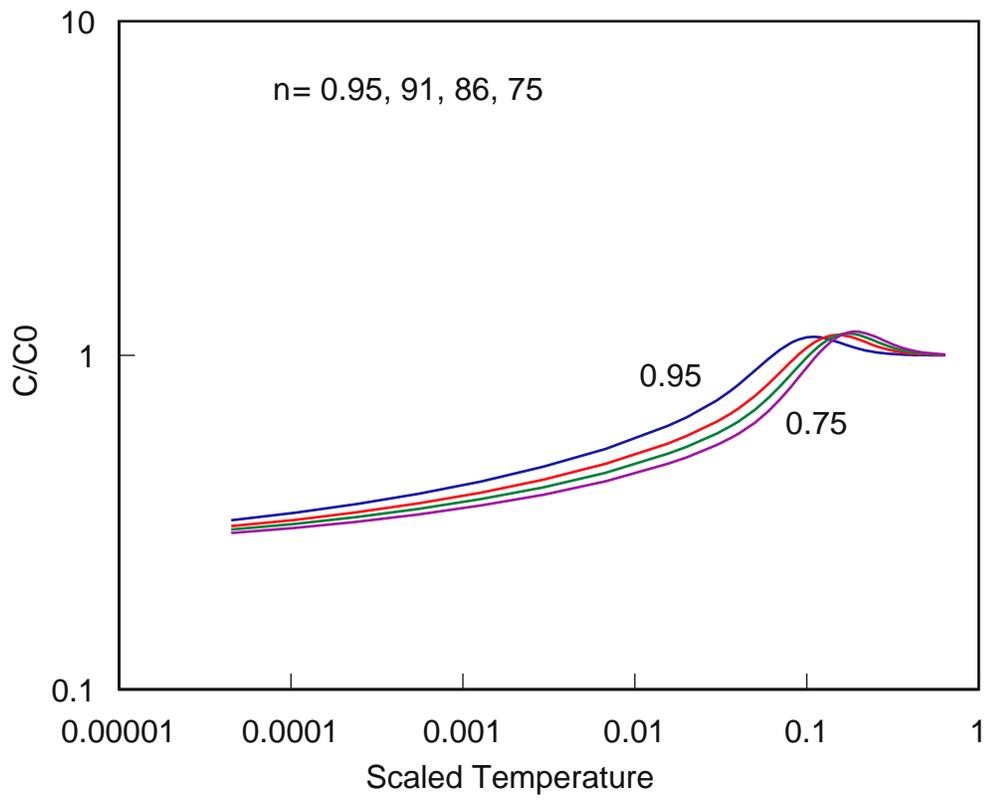
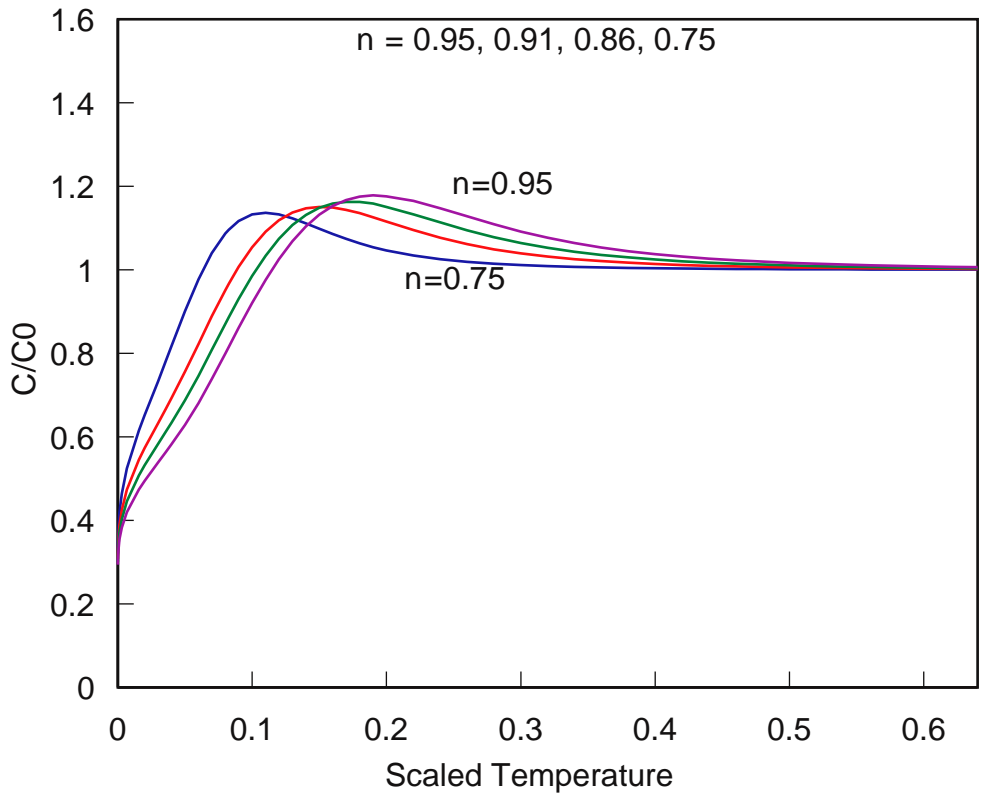
\end{document}